\documentclass[a4paper,twoside]{article}
\usepackage{epsfig}
\usepackage{subcaption}
\usepackage{calc}
\usepackage{amssymb}
\usepackage{amstext}
\usepackage{amsmath}
\usepackage{amsthm}
\usepackage{multicol}
\usepackage{pslatex}
\usepackage{algorithm2e}
\usepackage{booktabs}
\usepackage[bottom]{footmisc}
\usepackage{array}
\usepackage{multirow}
\usepackage{makecell}
\usepackage{wrapfig}
\usepackage{url} 

\usepackage{SCITEPRESS} 

\begin{document}

\title{An extreme Gradient Boosting (XGBoost) Trees approach to Detect and Identify Unlawful Insider Trading (UIT) Transactions%
   \thanks{This manuscript was presented at the 14th International Conference on Data Science, Technology and Applications (DATA 2025), Bilbao, Spain, June 10--12, 2025. \protect\url{https://data.scitevents.org/Home.aspx?y=2025}.}
.
}

\author{\authorname{Krishna Neupane\sup{1}\orcidAuthor{0000-0003-3911-3988}, Igor Griva\sup{2}\orcidAuthor{0000-0002-2291-233X}}
\affiliation{\sup{1}George Mason University, Department of Computational and Data Science}
\affiliation{\sup{2}George Mason University, Department of Mathematical Science}
\email{\{kneupan, igriva\}@gmu.edu}
}


\abstract{Corporate insiders have control of material non-public preferential
information (MNPI). Occasionally, the insiders strategically bypass
legal and regulatory safeguards to exploit MNPI in their execution of
securities trading. Due to a large volume of  transactions a detection of
unlawful insider trading becomes an arduous task for humans to examine
and identify underlying patterns from the insider's behavior. On the
other hand, innovative machine learning architectures have shown
promising results for analyzing large-scale and complex data with hidden
patterns. One such popular technique is eXtreme Gradient Boosting
(XGBoost), the state-of-the-arts supervised classifier. We, hence, resort
to and apply XGBoost to alleviate challenges of identification and
detection of unlawful activities. The results demonstrate that XGBoost
can identify unlawful transactions with a high accuracy of 97 percent
and can provide ranking of the features that play the most important
role in detecting fraudulent activities.}

\pagestyle{plain}

\onecolumn 
\maketitle 
\normalsize 
\setcounter{footnote}{0} 
\vfill

\section{\uppercase{Introduction}}
\label{sec:introduction}

Corporate insiders\footnotemark[1], in their privileged roles, access
material non-public information (MNPI)\footnotemark[2]. While the
Securities Exchange Act of 1934, specifically Section
10b-5\footnotemark[3], prohibits utilizing this information for
financial gain, detecting violations is challenging. Insiders often
employ creative strategies to conceal their trading activities. These
unlawful trades often mimic routine transactions
\cite{CohenPomorski2012}, making them opaque and difficult to
identify using traditional, manually-engineered approaches. Therefore,
effectively uncovering hidden patterns of such activity within
voluminous transaction data requires innovative methodologies that have
demonstrated effectiveness
\cite{mayo2022statistical}, \cite{varol2017online}.

Historically, research on detecting unlawful insider trading (UIT) has often been grounded in economic theories and legal analysis.  Kyle's 1985 paper provided the first significant theoretical formulation for unlawful insider trading (UIT), applying information asymmetry—the phenomenon of unequal information causing market disequilibrium—through a dynamic model that examines how private information affects prices, market liquidity, and its value \cite{Kyle1985}.  Following this foundational work, other studies analyzing information asymmetry in insider trading include Seyhun's 1986 study \cite{SeyhunHNejat1986Ipco}, which investigated insider and outsider trading profits and the determinants of insiders' predictive ability using a large transaction dataset, discussing implications for market efficiency. Rozeff and Zaman's 1988 paper \cite{rozeff1988market} examined whether publicly available insider trading data allows outsiders to earn abnormal profits, finding the anomaly persists but is largely explained by size and earnings/price effects when considering transaction costs. Lin and Howe's 1990 paper \cite{LinHowe1990} examined insider trading profitability in the OTC/NASDAQ market, finding insiders show timing ability but high transaction costs preclude outside investors from earning abnormal profits by mimicking them, and identifying determinants of insider profits. Huddart, Hughes, and Levine's 2007 study \cite{HuddartStevenJ2007IAaC} investigated the relationship between insiders' trades and firms' information asymmetry, analyzing whether proxies for information asymmetry are associated with insider trading patterns as predicted by informed trading theories. Finally, Armstrong, Jagolinzer, and Pagach's 2012 paper \cite{armstrong2012corporate} examined the relationship between corporate governance and firms' information environments, finding that state antitakeover laws were associated with decreased information asymmetry and increased financial statement informativeness. 

Complementing studies on information asymmetry, scholars have also been motivated by the theory of liquidity preferences to study unlawful insider trading. Amihud and Mendelson's 1987 paper examined how trading mechanisms affect price behavior and return patterns, highlighting their impact on market liquidity \cite{amihud1987trading}. Easley et al.'s 1996 paper investigated how information-based trading affects spreads for different stocks, finding it contributes to observed differences in market liquidity \cite{easley1996liquidity}. Pagano and Steil's 1996 paper investigated whether greater transparency in trading systems enhances market liquidity by reducing trading costs for uninformed participants \cite{pagano1996transparency}. More directly linking insider trading to liquidity, Cao, Chen, and Shen's 2004 paper tested the hypothesis that insider trading impairs market liquidity, finding that significant insider trading around IPO lockup expirations had little negative effect on effective spreads and improved other liquidity measures \cite{cao2004does}.

In addition to economic perspectives, legal scholars have debated whether insider trading should be fully lawful versus unlawful. Bainbridge's 2022 paper \cite{bainbridge2022manne} critically examined the evolving legal standards applied by Delaware courts to controlling shareholder transactions, contending that increased skepticism leads to overregulation and proposing reforms to reduce costs and encourage investment. Manne's foundational 1966 work \cite{manne1966insider} reexamined the debate on insider trading's role, arguing that informed trading facilitates the timely transmission of valuable information to top managers and large shareholders, thus contributing to market efficiency.

In contrast to the pro-lawful stance, the opposing camp argues that insider trading impedes and erodes investor confidence and increases agency costs, with research supporting the need for regulation. Gangopadhyay et al.'s 2022 study \cite{gangopadhyay2022profits} found that opportunistic insider trading profits, particularly from purchases, significantly decreased following the enactment of the Dodd-Frank Act, suggesting regulation impacts strategic insider behavior. Cumming et al.'s 2011 paper \cite{cumming2011exchange} examined stock exchange trading rules concerning market manipulation, insider trading, and broker-agency conflict across countries and over time, finding that differences in these rules significantly affect market liquidity.

Detection methods derived from these domains typically rely on
explicitly stated functional relationships and limited sets of
covariates (e.g., volume, prices, returns, book-to-market, influence,
sentiment, and so on)
\cite{jacobs2015determinants}, \cite{fishman1995mandatory}, \cite{john1997market}, \cite{leamer1978specification}.
These traditional approaches struggle to capture the interactiveness and
non-linearities inherent in data, leading to potential model
misspecifications and limited discovery of complex empirical
irregularities. Furthermore, techniques often employed, such as
time-series forecasting are known for their lack of scalability with
increasing data volumes and can be prone to over-generalization when
evaluated on single train/test splits
\cite{hand2009forecasting}, \cite{anderson2007new}, \cite{ge2000deformable},  \cite{hamilton1989new}, \cite{rabiner1986introduction}, \cite{box1972some}.

Addressing the limitations of traditional methods and the need for
innovative approaches, machine learning (ML) techniques, particularly
classifiers, represent a promising avenue for detecting complex hidden
patterns indicative of UIT
\cite{sundarkumar2015novel}, \cite{louzada2012bagging}. In the context of UIT,
numerous studies have leveraged various classification methods to
identify potential UIT based on data from events, news, public
information releases, and transaction patterns
\cite{li2022identification}, \cite{rizvi2022unsupervised}, \cite{seth2020predictive}, \cite{islam2018mining}, \cite{goldberg2003nasd}.

Among the scalable and data-driven ML techniques successfully applied in
this domain are ensemble methods, such as Random Forest (RF) and
XGBoost. These methods are effective because they learn and discover
empirical regularities directly from data without requiring pre-defined
functional relationships. Both RF and XGBoost have demonstrated success
in detecting, identifying, and characterizing UIT. Specific studies
illustrate this success. For instance, Deng et
al. \cite{deng2021intelligent} implemented RF in the Chinese
Securities Market with 26 features, accurately classifying over 75
percent of UIT. Building upon this, Neupane et
al. \cite{neupane2024randomforest} extended the feature space to 110
features, achieving over 95 percent accuracy with RF. Related work has
also utilized XGBoost for this purpose, with an effort by Deng et
al. \cite{deng2019identification} reporting 85 percent accuracy.
Drawing on these consistent and promising results, the current study
utilizes XGBoost, leveraging its architectural design for parallel
computing and its iterative process of updating parameters to strengthen
weak learners by implicitly engaging every feature. This approach
fundamentally addresses the drawbacks of manual feature engineering,
such as mis-specifications and omitted features, by inherently handling
inter-dependencies, multi-dimensionalities, and non-linearities in data
\cite{malhotra2021hybrid}, \cite{hou2020replicating}, \cite{iskhakov2020machine}, \cite{camerer2019artificial}, \cite{fudenberg2019predicting}.

This study makes several contributions. First, the feature space for
XGBoost-based UIT detection is extended from 26 to 110 features to
assess the impact on accuracy. Second, the analysis is based on a
significantly larger number of transactions from the US Securities
market compared to previous work. Third, a simplified parameter search
technique is employed for improved efficiency over external optimization
methods. Fourth, using two ranking techniques, distinct features that
play prominent roles in identifying unlawful trading within a mixed set
of institutional, trade, and financial features are identified, with
results compared both with and without removing correlation between
features.

The manuscript is organized as follows. Section
\ref{sec-method-proposed-method} describes the methodology,
outlining the theory behind various used techniques, hyper-parameter
tuning, performance measures and feature selection criteria. Section
\ref{sec-analysis-experimental-setup} describes the experimental
settings. Section \ref{sec-analysis-data} includes data
description, classification results, and feature ranking. Section
\ref{sec-conclusions-future} discusses the results and provides
conclusions and possible future directions.

\section{\uppercase{Proposed Methodology}}\label{sec-method-proposed-method}

To detect UIT, the paper implement XGBoost, a method well-known for its
ability to capture complex nonlinear interactions in the data, which is
a basis for attaining high out-of-sample accuracy. Designed for speed
and efficient memory management, XGBoost has demonstrated superior
performance across diverse applications, including credit scoring
\cite{mushava2022novel}, fraud detection \cite{zhang2020customer},
consumer credit risk evaluation \cite{wang2022research}, DNA sequence
identification \cite{sang2020hmmpred}, and climate science
\cite{wang2022analysis}. Moreover, as an ensemble method, it aligns
with techniques considered effective for empirical work in economics
\cite{athey2019impact}. The approach taken in this study leverages
corporate governance, trade, and finance data for detecting UIT by
extending the application of XGBoost to this domain. The methodology
also integrates Principal Component Analysis with XGBoost for
comparative analysis. For comparison, the results are contrasted with
previous studies, specifically those by \cite{deng2021intelligent}, \cite{deng2019identification}, and \cite{neupane2024randomforest}. These represent the only publicly
available comparative studies in this area.

\subsection{eXtreme Gradient Boosting (XGBoost)}\label{sec-method-gradient-boosting}

XGBoost was proposed by \cite{chen2016xgboost}, which is
a highly scalable and powerful algorithm belonging to the gradient
boosting family. It implements a distributed gradient tree boosting
strategy, training the model by sequentially learning from multiple weak
classifiers and iteratively updates them to correct errors from
preceding steps, while also allowing for efficient memory management.
This iterative process combines the updated weak learners into a
powerful ensemble. In summary, XGBoost trains its model through this
iterative boosting process: It starts with an initial base prediction.
Then, in each step, it calculates the errors (residuals), constructs and
fits a new decision tree to predict these residuals, and adds the tree
to the ensemble to minimize loss. Predictions are updated, new residuals
calculated, and this sequence is repeated for a set number of
iterations. The final prediction combines the outputs from all trees.
Formally, consider a training dataset,
\(\mathbb{D} = {(x_i, y_i)}_{i=1}^n\), where \(n\) is the number of
instances (rows) and each instance (\(x_i \in \mathbb{R}^m\)) is a
vector of \(m\) features (columns), \(y_i \in \mathbb{R}\) represents
the label for the \(i\)-th instance (e.g., 1 for unlawful, 0 for
lawful). The predicted value, \(\hat{y}_i\) for the \(i\)-th instance
from an ensemble model comprising \(K\) decision trees is given by the
sum of the predictions from each tree as in Equation
\ref{eq:prediction}.

\begin{equation}
  \hat{y}_i = \sum_{k=1}^K f_k(x_i), \quad f_k \in \mathbf{F},
  \label{eq:prediction}
\end{equation}
where $f_k$ denotes the $k$-th decision tree and $\mathbf{F}$ is the functional space containing all possible decision trees. XGBoost aims to minimize a regularized objective function $Obj$ to learn the set of trees ${f_k}_{k=1}^K$. This objective function combines the training loss and a regularization term to control model complexity. The loss function $\ell(y_i, \hat{y}_i)$ measures the difference between the actual label ($y_i$) and the predicted value ($\hat{y}_i$) for a single instance. The total training loss over the dataset is the sum of individual instance losses given by Equation \ref{eq:loss}.

\begin{equation}
  L(\mathbf{y}, \hat{\mathbf{y}}) = \sum_{i=1}^n \ell(y_i, \hat{y}_i),
  \label{eq:loss}
\end{equation}

where $\mathbf{y}$ and $\hat{\mathbf{y}}$ are the vectors of actual and predicted labels for all $n$ instances, respectively. The loss function $\ell$ can be selected based on the task (e.g.log loss for classification). During training, XGBoost iteratively adds trees, optimizing the objective function with respect to the parameters of the new tree being added at each step. The regularization term $\Omega(f_k)$ for the $k$-th decision tree $f_k$ is calculated based on the tree's structure and leaf weights given by Equation \ref{eq:regularization}.

\begin{equation}
  \Omega(f_k) = \gamma T_k + \frac{1}{2} \lambda \sum_{j=1}^{T_k} w_{k,j}^2,
  \label{eq:regularization}
\end{equation}

where $T_k$ is the number of leaf nodes in the $k$-th tree, $w_{k,j}$ is the prediction weight of the $j$-th leaf in the $k$-th tree (with ($w_{k,j}^2$) being its square), ($\gamma$) is the $L1$ regularization term on the number of leaves, and ($\lambda$) is the $L2$ regularization term on the leaf weights. These terms control tree pruning and the magnitude of leaf weights, respectively. The overall regularized objective function that XGBoost minimizes is defined as the sum of the total training loss and a regularization term $\Omega$ that penalizes the complexity of the trees given by Equation \ref{eq:objective}. 

\begin{equation}
  Obj = \sum_{i=1}^n \ell(y_i, \hat{y}_i) + \sum_{k=1}^K \Omega(f_k),
  \label{eq:objective}
\end{equation}

\subsection{Parameter Tuning}\label{sec-method-parameter-turning}

Tuning hyperparameters is crucial for many ML techniques, and XGBoost is
no exception; it is essential for minimizing the objective function and
controlling overfitting. These parameters, which can be categorized into
regularization, pruning, and sampling, influence the overall prediction
errors. For \textbf{Regularization}, commonly used hyperparameters are
Learning rate (\(\eta\)) and L2 regularization (\(\lambda\)). \(\eta\)
controls the step size (shrinkage) applied to weights at each boosting
iteration. Smaller \(\eta\) values lead to more conservative models and
require more boosting rounds. \(\lambda\) applies penalty to the leaf
weights based on the sum of their squares. Increasing \(\lambda\) makes
the model more conservative. For \textbf{Pruning}, the Minimum split
improvement (\(\gamma\)) parameter is used. It acts as a regularization
parameter specifying the minimum loss reduction required to make a
split, thereby controlling tree complexity and preventing overfitting.
To reduce variance and improve generalization, \textbf{Sampling} is
applied to data instances or features for each tree or iteration. In
this study, \textbf{Column sub-sampling} and \textbf{Row sampling} are
employed. \textbf{Column sub-sampling} refers to the fraction of
features randomly sampled per tree or per level when building trees,
while \textbf{Row sampling} is the fraction of data instances randomly
sampled per tree or per round.

\subsection{Feature Importance}\label{sec-method-feature-importance}

XGBoost's built-in feature ranking, a key tool for model interpretation
and feature selection, is analogous to that of RF, as both techniques
commonly use the mean decrease in impurity (Gini Score) during training.
However, this method is known to have limitations, such as bias towards
correlated and high-cardinality features, and relies solely on training
data. To provide a more robust ranking, a second feature ranking
approach was implemented. This approach involves decorrelating features
using hierarchical clustering and subsequently ranking them based on
permutation importance scores. Permutation importance is often preferred
over MDI as it directly measures a feature's impact on model performance
on unseen data and is less susceptible to training-phase biases. This
second approach follows the methodology described by \cite{neupane2024randomforest}.

\subsection{Principal ComponentAnalysis}\label{sec-method-principal-component-analysis}

The analysis in this study employs Principal Component Analysis (PCA), a
classic unsupervised technique for data decorrelation and compression.
This method has demonstrated effectiveness in various applications,
notably in studies on UIT (\cite{deng2021intelligent},
\cite{neupane2024randomforest}), and the detailed methodology followed
is based on that described by \cite{neupane2024randomforest}.

\subsection{Performance Measure}\label{sec-method-performance-measure}

Model performance is evaluated using a \(2 \times 2\) confusion matrix
organized by actual and predicted classes, schematically represented by
Table \ref{tbl-confusionMatrixSchematic}. Assuming `Lawful' is the
positive class (+) and `Unlawful' is the negative class (-), the matrix
yields four outcomes: True Positives (TP) for correct positive
predictions, True Negatives (TN) for correct negative predictions, False
Positives (FP) for negative instances incorrectly predicted as positive,
and False Negatives (FN) for positive instances incorrectly predicted as
negative. From this matrix, metrics like overall accuracy (ACC) and
Precision (PRE) are calculated. ACC measures the total proportion of
correct classifications, and PRE (for the positive class) is the
proportion of predicted positives that are truly positive. \vspace{0pt}

\begin{table}
  \centering 
  \captionsetup{width=\linewidth} 
  \caption{\label{tbl-confusionMatrixSchematic}Organization of the 2 × 2 grid of confusion matrix used to measure state of lawfulness of insider trading transactions}
  \small
  \begin{tabular}{| >{\centering\arraybackslash}p{0.15\linewidth} | >{\centering\arraybackslash}p{0.25\linewidth} | >{\centering\arraybackslash}p{0.2\linewidth} | >{\centering\arraybackslash}p{0.2\linewidth} |}
    \hline
    \multicolumn{4}{|c|}{Predicted Label (PP+PN)} \\
    \hline
    \multirow{3}{0.15\linewidth}{Actual\\ Labels} & Total Population & Positive & Negative \\ \cline{2-4}
                                                            & Lawful - Positive & True Lawful & False Unlawful \\ \cline{2-4}
                                                            & Unlawful - Negative & False Lawful & True Unlawful \\
    \hline
  \end{tabular}
\end{table}

\section{\uppercase{Experimental Setup}}\label{sec-analysis-experimental-setup}

The experimental settings broadly replicate Neupane et
al.(\cite{neupane2024randomforest}). Data originates from SEC Form 4
filings, linked with CRSP and Compustat-CapitalIQ trade and finance data
via personid, cik, and companyid. Comprising 3984 fully labeled
transactions (1992 unlawful) with 110 dimensions per row, the merged
dataset was used alongside a 320-transaction subset for comparison. Each
dataset subset, balanced (0.5:0.5 ratio) and sub-divided by feature sets
(original vs.~PCA-integrated), was then split deterministically 80
percent (training): 20 percent (test) for analysis. Numerical features
\(X_i\) (\(i = 1, \ldots, n\)) were normalized using the \(z\)-score
method\footnote{The $z$-score transformation standardizes features to have a mean of 0 and standard deviation of 1, placing predictors on a common scale ($\frac{X_i- \mu}{\sigma}$) \cite{gelman2008scaling}.},
while categorical features were one-hot-encoded. Hyperparameters such as
\(\eta\), \(\gamma\), max depth, and sample rate were initialized in a
random search space, with tuning conducted over 5 iterations within
5-fold cross-validation. Feature rankings were derived from MDI (based
on training data) and permutation importance (applied to training and
test data), with the latter's flexibility allowing for the ranking of
test data even after feature decorrelation. Correlation was removed by
performing hierarchical clustering based on Spearman rank-order
correlation and selecting a representative feature from each cluster.
The experiment was performed with 100 repetitions using scikit-learn and
xgboost libraries, where each repetition involved randomly sampling
lawful transactions.

\section{\uppercase{Analysis and Results}}\label{sec-analysis-data}

This section reports and interprets the performance of the implemented
methods based on confusion matrix metrics, drawing upon the dataset
characteristics illustrated in Table \ref{tbl-transactionsCount}.
Performance metrics, averaged over 100 experiments (Table
\ref{tbl-xgbComparativeConfusionMatrixKrishnaResultsOneHundredExperiments}),
are presented. Hyperparameter tuning was performed to optimize model
performance, involving 5-fold cross-validation and 100 repetitions,
using AUC as the stopping criterion. This process optimized parameters
such as \(ntrees\), \(\eta\), \(max \ depth\), \(\gamma\), and
\(sample \ rate\): for instance, \(ntrees\) was typically optimized to
values around \(500\) to \(520\), \(max \ depth\) often favored values
around \(16\), and \(\eta\) was right around the default value of
\(0.03\). To compare the reported metrics, benchmark results from \cite{deng2019identification} are presented in Table
\ref{tbl-rfComparativeConfusionMatrixBenchMarkMethods}.

\begin{table}
  \centering 
  \captionsetup{width=\linewidth} 
  \caption{\label{tbl-transactionsCount}Distribution of balanced unlawful and randomly selected lawful transactions. The right-hand side shows a random subset of this data matching transaction counts from Deng et al. (2019). Example referenced from Neupane et al. (2024).}
  \tiny

  \begin{tabular}{@{} >{\centering\arraybackslash}p{0.18\linewidth} 
                     >{\centering\arraybackslash}p{0.08\linewidth}  
                     >{\centering\arraybackslash}p{0.08\linewidth}  
                     >{\centering\arraybackslash}p{0.08\linewidth}  
                  @{} >{\centering\arraybackslash}p{0.02\linewidth} 
                  @{} >{\centering\arraybackslash}p{0.08\linewidth}  
                     >{\centering\arraybackslash}p{0.08\linewidth}  
                     >{\centering\arraybackslash}p{0.08\linewidth}  
                  @{} >{\centering\arraybackslash}p{0.02\linewidth} @{}} 
  \hline
  & \multicolumn{3}{c}{All Trans.} &  
  & \multicolumn{3}{c}{Subset of Trans.} &  \\ 
  \cline{2-4} \cline{6-8}
  Label & Sell     & Pur.     & Total     &  & Sell        & Pur.       & Total       &  \\
  \cline{2-4} \cline{6-8}
  Lawful   & 405      & 1587          & 1992      &  & 27          & 133            & 160         &  \\
  Unlawful & 318      & 1674          & 1992      &  & 26          & 134            & 160         & \\
  \hline
  \end{tabular}
  \end{table}

  \subsection{Results of Classification of Insider Trading Transactions}\label{sec-analysis-Results-Classification-Transactions}

  Performance varies with transaction count, feature set size, and PCA
  integration. The benchmark method (XGBoost-NSGAII) achieves an accuracy
  of \(84.99\) percent (Table
  \ref{tbl-rfComparativeConfusionMatrixBenchMarkMethods}). In the
  implemented settings using \(320\) transactions, the average ACC is
  \(83.105\) (Table
  \ref{tbl-xgbComparativeConfusionMatrixKrishnaResultsOneHundredExperiments}),
  closely approaching the benchmark. Performance with \(320\) transactions
  improves significantly to \(89.24\) percent ACC when PCA is not
  integrated. Furthermore, utilizing the full \(3984\) transactions
  consistently leads to improved performance across all feature set sizes
  and PCA conditions. For instance, using the full dataset, ACC averages
  \(90.61\) percent, surpassing the benchmark.

  \begin{table}
    \centering 
    \captionsetup{width=\linewidth} 
    \caption{\label{tbl-rfComparativeConfusionMatrixBenchMarkMethods}Performance evaluation metrics for benchmark methods applied to UIT detection referenced from Deng et al. 2021, Deng et al. 2019}
    
    \tiny 
    
    \setlength{\tabcolsep}{3pt} 
    
    \begin{tabular}{@{} >{\centering\arraybackslash}p{0.12\linewidth} 
                       >{\centering\arraybackslash}p{0.07\linewidth}  
                       >{\centering\arraybackslash}p{0.07\linewidth}  
                       >{\centering\arraybackslash}p{0.07\linewidth}  
                       >{\centering\arraybackslash}p{0.09\linewidth}  
                       >{\centering\arraybackslash}p{0.09\linewidth}  
                       >{\centering\arraybackslash}p{0.09\linewidth}  
                       >{\centering\arraybackslash}p{0.09\linewidth}  
                       >{\centering\arraybackslash}p{0.09\linewidth} @{}} 
    \hline
    
    & \multicolumn{3}{c}{} 
    & \multicolumn{2}{c}{Random Forest*} 
    & \multicolumn{3}{c}{XGBoost†} \\ 
    \cline{5-6} \cline{7-9} 
    
    Label & ANN & SVM & Adaboost & No PCA & With PCA & Classic & GA & NSGA II \\
    \hline
    
    ACC & 69.57 & 75.33 & 74.75 & 79.01 & 77.15 & 77.88 & 81.77 & 84.99\\
    FNR & 19.21 & 21.42 & 26.62 & 21.97 & 20.14 & 22.70 & 16.43 & 13.47\\
    FPR & 34.07 & 27.75 & 24.42 & 19.57 & 25.48 & 21.56 & 20.10 & 16.31\\
    TNR & 65.93 & 72.75 & 75.58 & 80.43 & 74.52 & 78.44 & 83.69 & 83.69\\
    TPR & 80.79 & 78.58 & 73.38 & 78.03 & 79.86 & 77.30 & 83.57 & 86.53\\
    PRE & - & - & - & - & - & 78.94 & - & - \\
    \hline 
    
    \end{tabular}
    
    \begin{center} 
      \begin{minipage}{0.99\linewidth} 
      \small 
      Notes:
      * \cite{deng2019identification},
      † \cite{deng2021intelligent}
      \end{minipage}
    \end{center}
    
    \end{table}

\vspace{1pt}

\begin{table}
  \centering 
  \captionsetup{width=\linewidth} 
  \caption{\label{tbl-xgbComparativeConfusionMatrixKrishnaResultsOneHundredExperiments}Average of the performance metrics of 100 experiments in 5-fold cross-validation. The first four columns are based on 320 random selections from 3984 transactions matching the count of the previous study.}
  
  \tiny 
  
  \setlength{\tabcolsep}{2pt} 
  
  \begin{tabular}{@{} >{\raggedright\arraybackslash}p{0.1\linewidth} 
                     >{\raggedleft\arraybackslash}p{0.09\linewidth}  
                     >{\raggedleft\arraybackslash}p{0.09\linewidth}  
                     >{\raggedleft\arraybackslash}p{0.09\linewidth}  
                     >{\raggedleft\arraybackslash}p{0.09\linewidth}  
                  @{\hspace{1pt}} 
                     >{\raggedleft\arraybackslash}p{0.09\linewidth}  
                     >{\raggedleft\arraybackslash}p{0.09\linewidth}  
                     >{\raggedleft\arraybackslash}p{0.09\linewidth}  
                     >{\raggedleft\arraybackslash}p{0.09\linewidth} @{}} 
  \hline 
  
  & \multicolumn{4}{c}{\thead{\tiny Subset (n=320)}} 
  & \multicolumn{4}{c}{\thead{\tiny 3984 Trans.}} \\ 
  \cmidrule(lr){2-5} \cmidrule(lr){6-9} 
  
  Metric & \multicolumn{2}{c}{25 Features} 
         & \multicolumn{2}{c}{110 Features} 
         & \multicolumn{2}{c}{25 Features} 
         & \multicolumn{2}{c}{110 Features} \\ 
  \cmidrule(lr){2-3} \cmidrule(lr){4-5} \cmidrule(lr){6-7} \cmidrule(lr){8-9}
  
  & No PCA & With PCA & No PCA & With PCA & No PCA & With PCA & No PCA & With PCA \\
  
  ACC & 83.34 & 78.79 & 89.24 & 81.05 & 98.12 & 97.43 & 99.02 & 97.96\\
  PRE & 84.67 & 79.38 & 89.59 & 80.01 & 98.19 & 97.01 & 97.32 & 97.32 \\
  TPR & 81.88 & 78.7 & 89.3 & 83.5 & 98.05 & 97.87 & 98.98 & 98.64\\
  FNR & 18.12 & 21.3 & 10.7 & 16.5 & 1.95 & 2.13 & 1.02 & 1.36\\
  FPR & 15.2 & 21.12 & 10.82 & 21.39 & 1.8 & 3.01 & 0.93 & 2.71\\
  TNR & 84.8 & 78.88 & 89.18 & 78.61 & 98.2 & 96.99 & 99.07 & 97.29\\
  \hline 
  
  \end{tabular}
  
  \end{table}
 
 Based on the data illustrated in Table \ref{tbl-transactionsCount}, the
 performance of implemented methods is compiled and presented in
 confusion matrix metrics, averaged over 100 experiments and 5-fold
 cross-validation (Table
 \ref{tbl-xgbComparativeConfusionMatrixKrishnaResultsOneHundredExperiments}).
 To compare the results, those from \cite{deng2021intelligent} and \cite{deng2019identification} are
 compiled in Table
 \ref{tbl-rfComparativeConfusionMatrixBenchMarkMethods}. Among the
 benchmarks, XGBoost-NSGAII (last column of Table
 \ref{tbl-rfComparativeConfusionMatrixBenchMarkMethods}) achieves an
 accuracy of 84.99 percent, the highest. Comparatively, in Table
 \ref{tbl-xgbComparativeConfusionMatrixKrishnaResultsOneHundredExperiments}
 that uses 320 transactions, the ACC with 25 features is 83.34 (first
 column of Table
 \ref{tbl-xgbComparativeConfusionMatrixKrishnaResultsOneHundredExperiments}),
 a very close result compared to the benchmark. The performance declines
 to 78.79 percent when PCA is used in the same setting. But with the
 addition of features (110 Features) within the same settings (320
 transactions), the results start approaching the benchmark's highest
 performance. As the number of transactions is added (3984 transactions),
 with either limited set of features (25) or additional (110), the ACC
 starts improving substantially. A notable performance increase,
 averaging 90.61 percent ACC, is observed when using the full 3984
 transactions with either 25 or 110 features (with or without PCA).
 
 Beyond overall accuracy, other key metrics from the confusion matrix
 provide further insights into performance (Table
 \ref{tbl-xgbComparativeConfusionMatrixKrishnaResultsOneHundredExperiments}).
 For metrics where higher values indicate better performance -- True
 Positive Rate (TPR), True Negative Rate (TNR), and Precision (PRE) --
 the implemented method generally demonstrates competitive or superior
 results compared to benchmark methods (Table
 \ref{tbl-rfComparativeConfusionMatrixBenchMarkMethods}). While the
 benchmark's best reported TPR, TNR, and PRE are \(86.53\) percent,
 \(83.69\) percent, and \(78.94\) percent respectively, the implemented
 method achieves significantly higher values in several scenarios (Table
 \ref{tbl-xgbComparativeConfusionMatrixKrishnaResultsOneHundredExperiments}).
 For instance, using all \(3984\) transactions, TPR averages
 approximately \(98.38\) percent (reaching a high of \(98.98\) percent),
 and TNR averages approximately \(97.9\) percent (reaching a high of
 \(99.07\) percent). Consistent with ACC, TPR and TNR improve with
 increased data size.
 
 Conversely, for metrics where lower values indicate better performance
 -- False Positive Rate (FPR) and False Negative Rate (FNR) -- the
 implemented method also shows strong results, particularly with
 increased data (Table
 \ref{tbl-xgbComparativeConfusionMatrixKrishnaResultsOneHundredExperiments}).
 Compared to benchmark FPRs which average \(16.31\) percent (Table
 \ref{tbl-rfComparativeConfusionMatrixBenchMarkMethods}), the implemented
 method's FPR averages \(17.13\) percent with \(320\) transactions but
 drops significantly to approximately \(2.11\) percent with \(3984\)
 transactions. Similarly, benchmark FNRs range from \(13.47\) percent to
 \(26.62\) percent (Table
 \ref{tbl-rfComparativeConfusionMatrixBenchMarkMethods}), whereas the
 implemented method's FNR sees a substantial reduction from \(16.66\)
 percent with \(320\) transactions to a remarkable \(1.62\) percent with
 \(3984\) transactions, highlighting few missed unlawful transactions
 with more data. The impact of PCA varies; on average, performance
 metrics are better when PCA is not integrated.
 
 A direct comparison was made between the performance metrics of the
 implemented XGBoost method (Table
 \ref{tbl-xgbComparativeConfusionMatrixKrishnaResultsOneHundredExperiments})
 and the Random Forest results from Table 5 of \cite{neupane2024randomforest}, who used the same experimental
 conditions. Both models achieved exceptionally high performance when
 trained and evaluated on the full set of \(3984\) transactions,
 demonstrating strong accuracy and low error rates across various
 configurations (\(25/110\) features, with/without PCA). A detailed
 comparison highlights key strengths of the implemented XGBoost method.
 XGBoost achieves accuracy exceeding \(99\) percent in optimal
 configurations (Table
 \ref{tbl-xgbComparativeConfusionMatrixKrishnaResultsOneHundredExperiments})
 and demonstrates remarkably strong control over false negative rates,
 reaching a minimum FNR of \(1.02\) percent, which is marginally lower
 than the best Random Forest FNR (\(1.07\) percent) reported in Table 5 of \cite{neupane2024randomforest} . This strong performance in minimizing missed unlawful
 transactions, alongside high overall accuracy and robust control over
 other error rates, positions XGBoost as a highly effective and
 potentially preferred classifier for this task.

 \subsection{Variable Importance}\label{sec-analysis-variable-importance}

 
   
 
 
 The strong performance achieved by XGBoost (see Table
 \ref{tbl-xgbComparativeConfusionMatrixKrishnaResultsOneHundredExperiments}),
 which consistently outperformed benchmark studies, warrants an
 investigation into the contributions of individual input features to UIT
 classification. Analyzing these contributions enhances model
 explainability and interpretability. Therefore, to address this common
 limitation of many ML methods, feature importance ranking was conducted
 using XGBoost's inbuilt Mean Decrease of Impurity (based on Gini
 Scores), a training-data-based technique influenced by correlation, and
 permutation importance, a computationally expensive method that can be
 applied to training and test data after decorrelation using hierarchical
 clustering and representative feature selection (see Section
 \ref{sec-method-feature-importance} for details).

 \begin{figure}[hbt!] 
  \centering
  \includegraphics[width=1.0\linewidth]{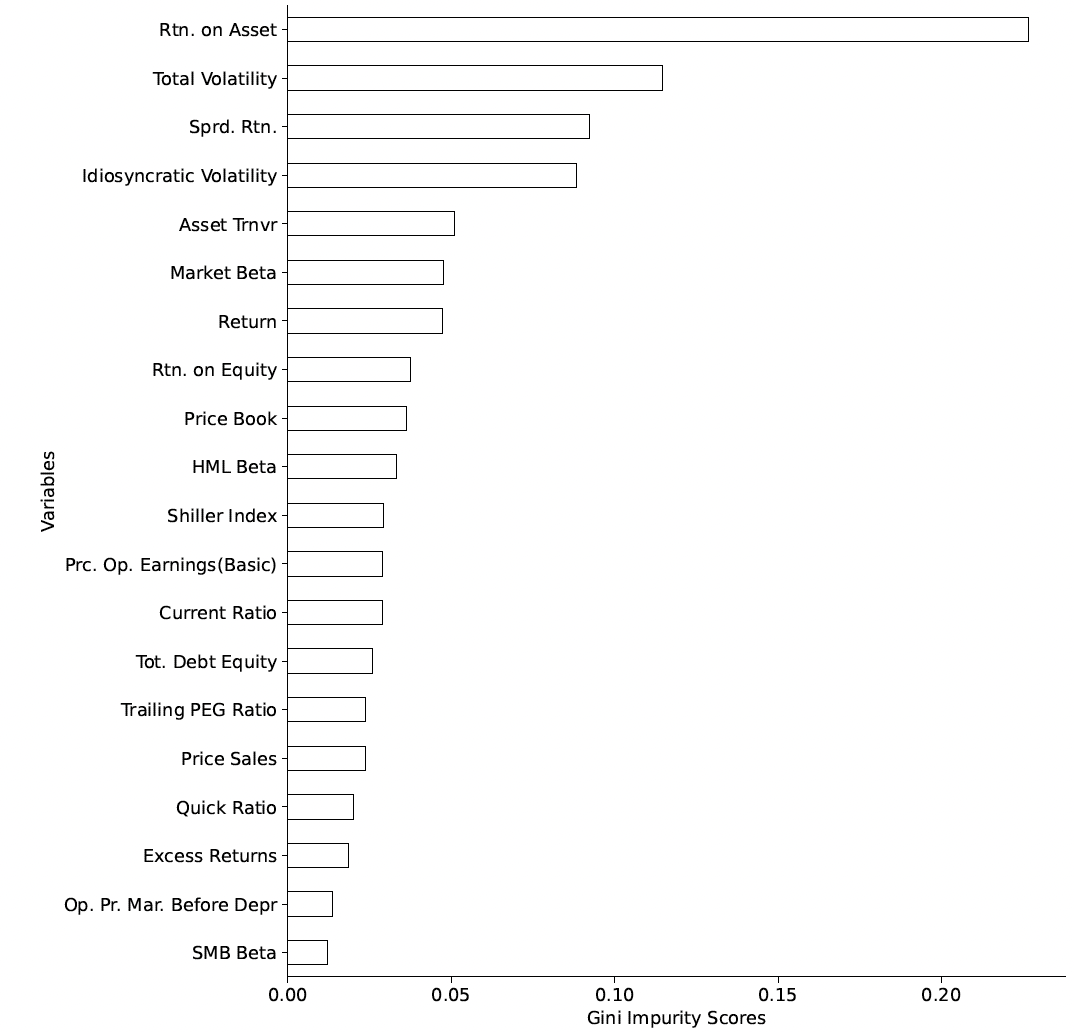} 

  \caption{Ranking of the importance of features based on Mean Decrease in Impurity extracted during training phase (see \cite{neupane2024randomforest} for details)}
  \label{fig:featureImportance_MDI} 
\end{figure}
 
\begin{figure}[hbt!] 
  \centering
  \includegraphics[width=1.0\linewidth]{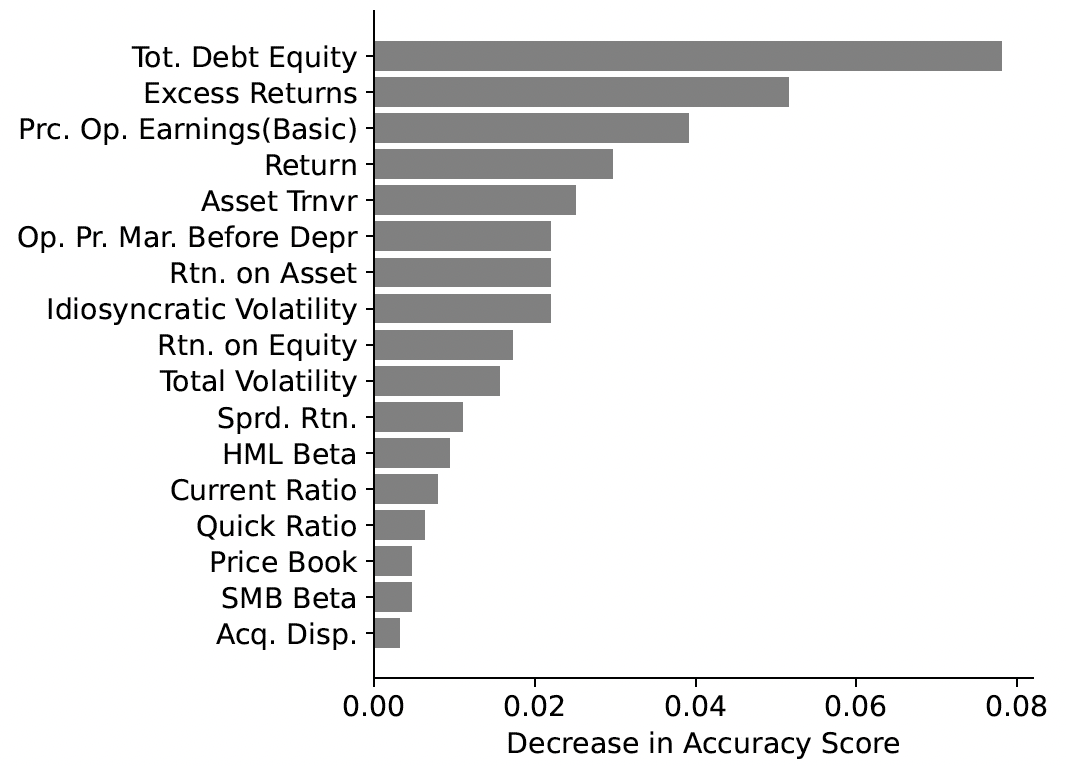} 

  \caption{Ranking of the importance of features based on Permutation Importance (see \cite{neupane2024randomforest} for details)}
  \label{fig:featureImportance_before_decorrelation_permImp}
\end{figure}

 Figures \ref{fig:featureImportance_MDI} and
 \ref{fig:featureImportance_before_decorrelation_permImp} are horizontal
 bar charts illustrating feature importance rankings, where the length of
 each bar indicates the importance score of a specific feature, with
 features ordered from most important at the top to least important at
 the bottom. A longer bar signifies higher importance according to the
 specific method used. Figure \ref{fig:featureImportance_MDI} presents
 features ranked by MDI scores, while Figure
 \ref{fig:featureImportance_before_decorrelation_permImp} displays the
 ranking obtained using Permutation Importance before applying
 decorrelation. As discussed, MDI-based ranking is based solely on
 training data and is known to be particularly sensitive to highly
 correlated features, which are common in financial datasets, potentially
 not generalizing well to test samples
 (\cite{meinshausen2008hierarchical}). Permutation importance is
 employed to address these shortcomings. This model-agnostic method
 evaluates feature contribution by measuring the decrease in model
 performance when a feature's values are randomly permuted
 (\cite{nembrini2018revival}), and importantly, can be applied to test
 data, unlike MDI.

 However, a visual comparison of Figure \ref{fig:featureImportance_MDI}
 and Figure \ref{fig:featureImportance_before_decorrelation_permImp}
 reveals notable differences in the top-ranked features and their
 relative importance. While MDI tends to rank profitability and
 volatility-related features highly (e.g., Return on Asset, Total
 Volatility), Permutation Importance before decorrelation ranks features
 such as Total Debt to Equity, Excess Returns, and Price Operating
 Earnings (Basic) as most important. This discrepancy, highlights that
 Permutation Importance is also significantly affected by correlation
 when applied to correlated data. In highly correlated datasets,
 permuting one feature might not significantly decrease performance if a
 highly correlated feature provides redundant information to the model.
 Consequently, neither the MDI ranking nor the Permutation Importance
 ranking before decorrelation provides a fully reliable measure of true
 feature importance in this highly correlated financial dataset. This
 underscores the importance of applying permutation importance after
 decorrelation for a more accurate assessment.
 
 \begin{figure}[hbt!] 
  \centering 
  \includegraphics[width=1.0\linewidth]{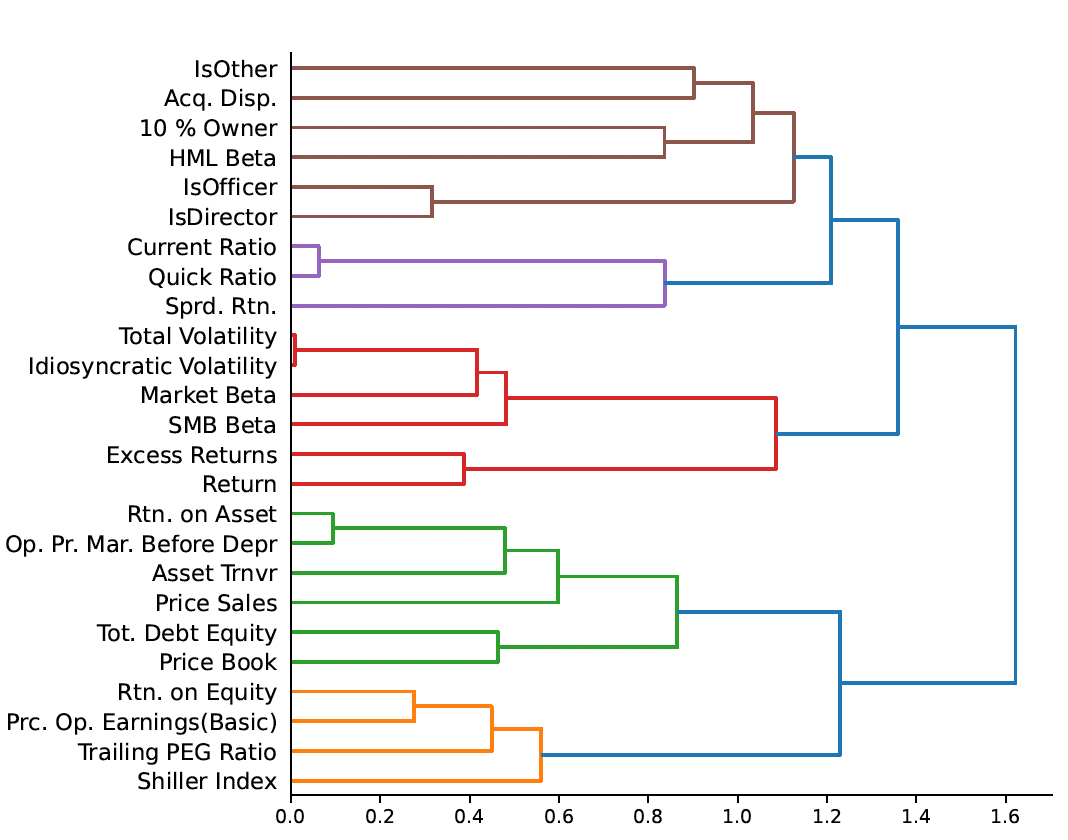} 

  \caption{Hierarchical clustering of features using Spearman rank-order correlations visualized by this dendrogram, showing the relationships and grouping of features based on similarity.}
  \label{fig:fig_2_hierarchial_clust} 
\end{figure}


 
\begin{figure}[hbt!] 
  \centering 
  \includegraphics[width=0.9\linewidth]{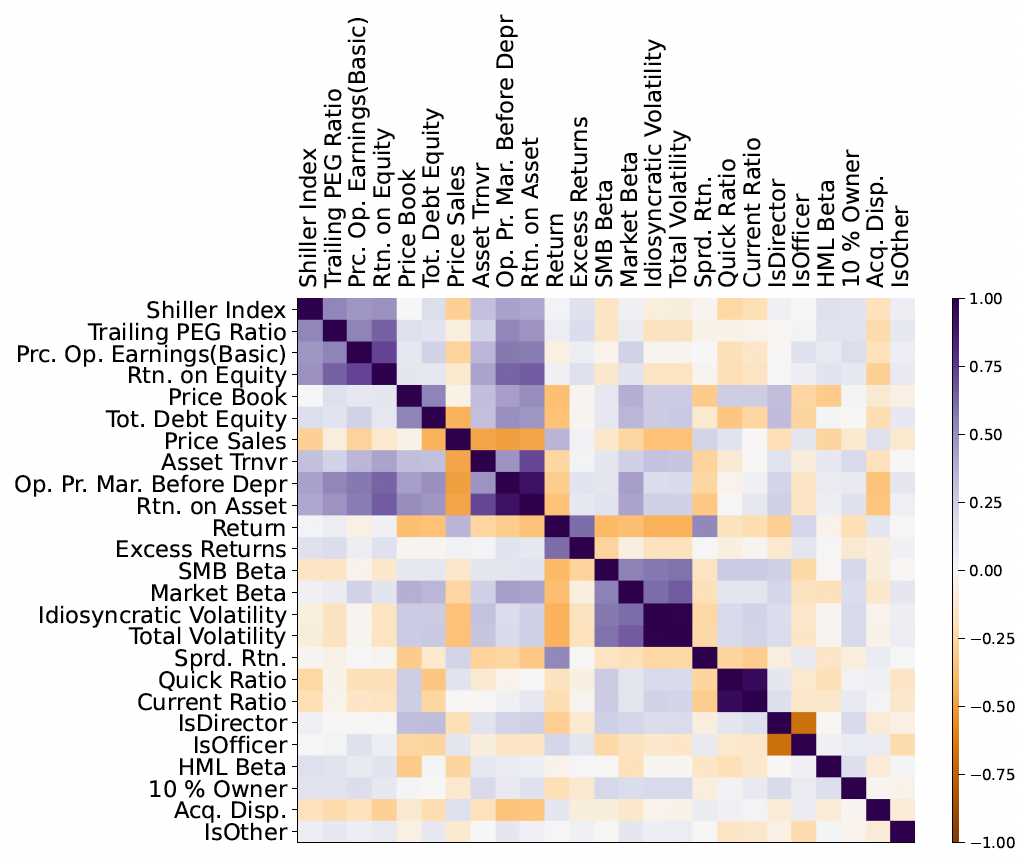} 

  \caption{Spearman Rank-Order Correlation Matrix for Selected Features (Illustrative), visualizing pairwise correlations to aid in identifying groups. In the color gradient, dark purple represents (perfect positive correlation), and dark orange represents (perfect negative correlation).}
  \label{fig:fig_3_corrl_mat} 
\end{figure}

To mitigate the impact of correlation on feature ranking, hierarchical
clustering was performed based on Spearman rank correlation, using
Ward's minimum variance linkage and a distance matrix derived from the
correlation matrix. This process is visualized in Figure
\ref{fig:fig_2_hierarchial_clust}, which shows the resulting dendrogram
and Figure \ref{fig:fig_3_corrl_mat}, which displays the correlation
matrix as a heatmap. In the heatmap (Figure \ref{fig:fig_3_corrl_mat}),
features are arranged along both axes, and the color intensity of each
cell indicates the strength of the correlation between the corresponding
features, with darker colors representing stronger positive or negative
correlations; the diagonal shows perfect correlation of each feature
with itself. The dendrogram (Figure \ref{fig:fig_2_hierarchial_clust})
illustrates the hierarchical clustering results; the vertical branches
show how features are merged into clusters based on their distance
(indicated on the horizontal axis), with shorter branches connecting
more similar features. Features grouped by branches form \emph{clades}.
For instance, Price Earnings (basic) and Return on Equity form a clade,
connected together with the Trailing PEG Ratio, forming the leftmost
clade. A representative feature was then selected from each cluster
based on these relationships.

\begin{figure}[hbt!] 
  \centering 
  \includegraphics[width=1.0\linewidth]{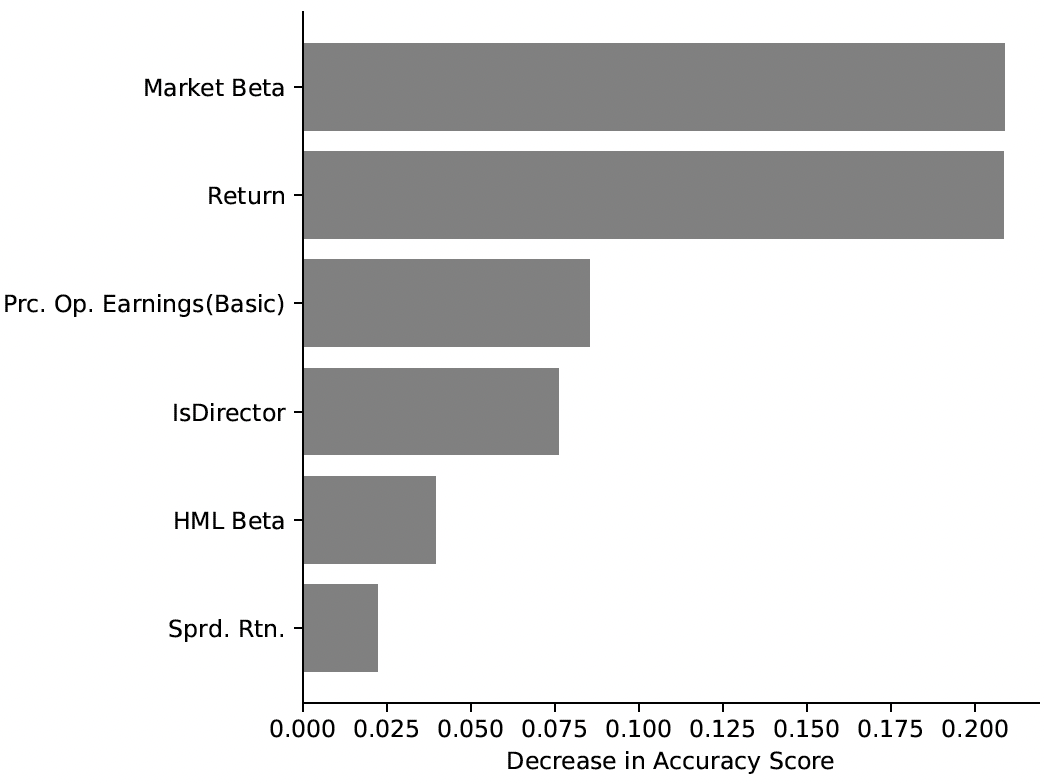} 

  \caption{Ranking of feature importance based on permutation values after removal of correlation due to hierarchical clustering. The horizontal axis is the scaled value of relative importance. The vertical axis represents the variables. The bars are organized in descending order of the relative importance.}
  \label{fig:fig_5_perm_imp_after_decorrln} 
\end{figure}

Figure \ref{fig:fig_5_perm_imp_after_decorrln} illustrate the impact of
correlation removal on feature ranking. Figure
\ref{fig:featureImportance_before_decorrelation_permImp} shows the
ranking obtained using Permutation Importance before decorrelation,
while Figure \ref{fig:fig_5_perm_imp_after_decorrln} displays the
ranking after hierarchical clustering and representative feature
selection. The ranking in Figure \ref{fig:fig_5_perm_imp_after_decorrln}
highlights the prominence of features related to market risk, corporate
governance, and valuation. Prominent features include Market \(\beta\),
Return, Price Operating Earnings (Basic), and IsDirector. Compared to
the ranking before decorrelation (Figure
\ref{fig:featureImportance_before_decorrelation_permImp}), the analysis
after decorrelation emphasizes features such as Market \(\beta\) and
IsDirector, which hold higher ranks. Price Operating Earnings (Basic)
also appears more influential after decorrelation, consistent with its
role as an important gauge for company valuation. The high ranking of
IsDirector indicates the importance of a role on the company's board in
influencing UIT. The importance of market \(\beta\) and value premium
features (like HML \(\beta\)) in this decorrelated context aligns with
financial theories, particularly considering the potential institutional
influence of executives on policies (e.g., dividend policy,
\cite{campbell1988dividend}, \cite{GrinblattMarkS1984Tveo}).

A comparative assessment of feature importance rankings from MDI (Figure
\ref{fig:featureImportance_MDI}), Permutation Importance before
decorrelation (Figure
\ref{fig:featureImportance_before_decorrelation_permImp}), and
Permutation Importance after decorrelation (Figure
\ref{fig:fig_5_perm_imp_after_decorrln}) reveals significant differences
across the three approaches. While MDI and Permutation Importance
applied before decorrelation produce differing rankings across the full
feature set, both methods are substantially affected by the presence of
highly correlated features common in financial data, leading to
potentially misleading importance scores. In contrast, the Permutation
Importance ranking after hierarchical clustering and representative
feature selection (Figure \ref{fig:fig_5_perm_imp_after_decorrln}) shows
a distinct set of prominent features and generally higher importance
scores for a subset of representatives. Following decorrelation,
features such as Market \(\beta\), Return, Price Operating Earnings
(Basic), and IsDirector emerge as highly influential in Figure
\ref{fig:fig_5_perm_imp_after_decorrln}. Results are consistent with
previous studies; the top features contributing most to the prediction
of unlawful activities are related to ownership, influence, and market
risk, indicating that daily activities in the capital market play an
important role in determining UIT. The disparity among the three
rankings underscores the profound impact of correlation on feature
importance measures and highlights why the ranking obtained after
decorrelation (Figure \ref{fig:fig_5_perm_imp_after_decorrln}) provides
a more reliable understanding of true feature contributions by
mitigating the masking effects of correlation.

In summary, the overall results of the supervised classifier, presented
in Table
\ref{tbl-xgbComparativeConfusionMatrixKrishnaResultsOneHundredExperiments},
demonstrate strong performance. To note, the classifier demonstrated a
consistent performance with high true positive and negatives as well as
a low false positive rate (fall out rate). This is crucial, as
wrongfully classifying an unlawful transaction as lawful is anecdotally
equivalent to a courtroom acquittal. The obtained false positive rates,
as shown in Table
\ref{tbl-xgbComparativeConfusionMatrixKrishnaResultsOneHundredExperiments},
compare favorably against benchmark methods presented in Table
\ref{tbl-rfComparativeConfusionMatrixBenchMarkMethods}, showing the
model successfully minimizes false alarms. Furthermore, XGBoost
demonstrates a thorough examination of information, leading to low false
negative rates (miss rate), as evident in Table
\ref{tbl-xgbComparativeConfusionMatrixKrishnaResultsOneHundredExperiments}.
Just as an incorrect incarceration has high stakes, misclassifying a
lawful transaction as unlawful is critical. The results indicate XGBoost
does not disregard or overlook hidden information, resulting in low
missing rates. In addition to controlling false classifications, the
proposed method produces strong true positive results, correctly
identifying lawful transactions. The high ratio of true negative to
negative further confirms the model's ability to correctly identify
unlawful transactions as unlawful. XGBoost effectively handles both
lawful and unlawful transactions across different scenarios, even when
the unlawful transactions are unchanged and lawful ones are randomly
sampled (\(50\) percent). The simple parameter tuning method proved to
be an effective strategy for achieving high accuracy. Finally, the
analysis indicates that decorrelation is impactful; by decorrelating,
corporate and institutional features like IsDirector gained prominence
in the ranking, appearing alongside key trade and finance features
(\cite{sigrist2023comparison}, \cite{meinshausen2008hierarchical}).

\section{\uppercase{Conclusions and Future Work}}\label{sec-conclusions-future}

In a high-dimensional feature space approach shows an excellent
performance to detect the UIT with accuracy over 97 percent. The
reliability of the results is assured by averaging them from 5-fold
cross-validation. The experiments run 100 times with a new set of lawful
transactions randomly sampled from a pool of 9.6 millions. Overall,
comparing the implemented XGBoost results (Table
\ref{tbl-xgbComparativeConfusionMatrixKrishnaResultsOneHundredExperiments})
with Random Forest results (Table 5 of \cite{neupane2024randomforest}) and the benchmark methods (Table
\ref{tbl-rfComparativeConfusionMatrixBenchMarkMethods}), the implemented
XGBoost method demonstrates high performance for UIT detection,
comparing favorably against the other methods, notably achieving higher
overall accuracy and remarkably lower false negative rates. Besides, the
results demonstrate that XGBoost provides the ranking of the features
that play the most important role in identification of the UIT. Those
features related to governance, financial and trading can be manipulated
by the corporate insiders for personal unlawful financial gains and
naturally contribute to uncovering fraudulent behaviors. Therefore, the
application of the advanced supervised machine learning techniques may
have significant practical impact on automated detection of the UIT.

For the future, the credibility of the detection of UIT can be improved
with the help of causality analysis. \cite{athey2019impact}) emphasizes decision trees are the most relevant
machine learning techniques to extract underlying causality. As a domain
agnostic, an effective decision trees method designed to handle large
datasets, XGBoost is a promising candidate for the future explorations.
Exploring XGBoost-causality nexus therefore may provide a high-stake
end-to-end utility and transparency to the SEC's overall process related
to the detection of insider trading. Researchers, further, can
contribute by studying the relationship between
classification-causality. Besides, tying features to an economic, a
financial and/or an institutional theory reduces the uncertainty and
inexplainability of models \cite{harvey2016and}. Therefore,
implementing decision tree methods to explain the tenets of UIT within
the realm of the economic and/or financial theories that includes
features analyzed in this research (\(25\) or \(110\)) or \(447\) as
proposed by \cite{hou2020replicating} is a valuable future
direction. In addition, during the experiments the random grid-search of
the hyper-parameters with a preset of the lower and upper-bound was
implemented that which may potentially warrant resource waste with
growing features space. In the future, by exposing and comparing results
from the alternative parameter optimization techniques, such as,
Bayesian Optimization, Grid Search, Evolutionary and so on is another
avenue to follow. Further, apart from the one-hot encoding method
applied to encode categorical features, meaningful insights can be
extracted by exploiting the existing relationships with application of
more advanced methods (e.g, target embedding)
\cite{rodriguez2018beyond} which remains unexplored in the context of
UIT to the best of current knowledge.

\bibliographystyle{apalike}
{\small
\bibliography{example}}

\end{document}